\documentclass[aps,floatfix,twocolumn,showpacs]{revtex4}
\usepackage{natbib}
\usepackage{dcolumn}
\usepackage{graphicx}
\newcommand{\physrep}{Phys.~Rep.}

\begin{document}
\title{The Contribution of the Cosmological Constant to the Relativistic Bending of Light Revisited}
\author{Wolfgang Rindler}
\author{Mustapha Ishak\footnote{Electronic address: mishak@utdallas.edu}}
\affiliation{
Department of Physics, The University of Texas at Dallas, Richardson, TX 75083, USA}
\date{\today}
\begin{abstract}
We study the effect of the cosmological constant $\Lambda$ on the bending of light by a concentrated spherically symmetric mass. Contrarily to previous claims, we show that when the Schwarzschild-de Sitter geometry is taken into account, $\Lambda$ does indeed contribute to the bending.
\end{abstract}
\pacs{95.30.Sf,98.80.Es,98.62.Sb}
\maketitle
\section{introduction}
In the ongoing effort to understand the nature of the dark energy that would be responsible for accelerating the expansion of our universe, it is of interest to investigate how the various candidates for this role differ in other observable effects. Review papers on the cosmic acceleration problem and the dark energy associated with it can be found, {\it e.g.}, in \cite{review1,review2,review3,review4,review5,review6,review7,review8} and references therein. One of the prime candidates is the cosmological constant $\Lambda$. From this point of view, various authors have studied the possible contribution of $\Lambda$ to the local bending of light. Even though such an effect would be many orders of magnitude too small to be measured with presently available instruments, it might in principle constitute one of the distinguishing characteristics of $\Lambda$. Yet there seems to be a generally uncontested perception in the literature that the basic light-bending effect of a concentrated spherically symmetric mass, as originally obtained by Einstein, is the same - by a fortuitous canceling of terms - whether or not one includes a cosmological $\Lambda$ term in the general-relativistic field equations. The purpose of our paper is to prove the contrary

The argument for the non-influence of $\Lambda$ was apparently first made in \cite{Islam} and has been re-made and re-affirmed by other authors, see for example \cite{Freire,Kerr,Kagramanova,Finelli,Sereno}. The common basis of their arguments is that, in the Schwarzschild-de Sitter metric (first derived by Kottler \cite{Kottler}), which applies when $\Lambda$ is included, $\Lambda$ nevertheless drops out of the exact $r,\phi$ differential equation for a light path (null geodesic). Hence also the integrated orbital $r,\phi$ relation for a light path is the same with or without $\Lambda$. And we agree with that.

But the differential equation and its integral are only half the story. The other half is the metric itself, which determines the actual observations that can be made on the $r,\phi$ orbit equation. When that is taken into account, a quite different picture emerges: $\Lambda$ {\it does} contribute to the observed bending of light! In fact, as intuition would suggest, since a positive $\Lambda$ effectively counteracts gravity, a positive $\Lambda$ diminishes the classical Einstein bending of light, as we shall show.

\section{The geometry and the bending of light}

The metric we shall be concerned with here (as were the other authors) is the above-mentioned Schwarzschild-de Sitter (SdS) metric
\begin{equation}
ds^2=\alpha(r) dt^2 - \alpha(r)^{-1} dr^2-r^2 (d\theta^2+sin^2(\theta) d\phi^2)
\label{eq:metric}
\end{equation}
where
\begin{equation}
\alpha(r) \equiv 1-\frac{2m}{r}-\frac{\Lambda r^2}{3},
\label{eq:alpha}
\end{equation}
and where, in the presently used relativistic units ($c=G=1$), $m$ is the mass of the central object. In the limiting case $\Lambda=0$ the metric (\ref{eq:metric}) reduces to the Schwarzschild metric; in the other limiting case, $m=0$, it reduces to the static form of the metric of de Sitter spacetime.

As is well known (e.g. Eq. (11.18) in \cite{Rindler}), the spatial equatorial coordinate "plane" $\theta=\pi/2$ (like all other such central "planes") in Schwarzschild spacetime, having 2-metric
\begin{equation}
dl^2=\Big{(}1-\frac{2m}{r}\Big{)}^{-1}dr^2+r^2 d\phi^2,
\label{Sch2metric}
\end{equation}
has an intrinsic geometry identical to that of the so-called Flamm paraboloid of revolution \cite{Flamm}, whose equation is 
\begin{equation}
z^2=8 m (r-2m)
\end{equation}
in Euclidean 3-space referred to cylindrical polar coordinates ($r,\phi,z$). (It is the surface labeled $\Sigma^3$ in our Figure 1.) The importance of the "central planes" lies in the fact that every orbit -by symmetry- lies in one of them.

The central planes of SdS spacetime naturally have a different intrinsic geometry. Theirs is determined by the 2-metric
\begin{equation}
dl^2=\Big{(}1-\frac{2m}{r}-\frac{\Lambda r^2}{3}\Big{)}^{-1}dr^2+r^2 d\phi^2.
\label{eq:SdS2metric}
\end{equation}
Near the central mass (where $m/r$ dominates over $\Lambda r^2$) we get essentially the Flamm geometry. Far from the central mass (where $\Lambda r^2$ dominates over $m/r$) we get the geometry of a sphere of radius $a=\sqrt{3/\Lambda}$:
\begin{equation}
dl^2=\Big{(}1-\frac{r^2}{a^2}\Big{)}^{-1}dr^2+r^2 d\phi^2.
\label{deSitter2metric}
\end{equation}
(See Eq. (16.18) in \cite{Rindler} with $r=a \eta$, and Figure 16.3 there, where $\eta$ corresponds to the de Sitter $r$.)

Hence the surface of revolution that replaces the Flamm paraboloid in the case of SdS spacetime is a combination of a Flamm paraboloid and a sphere, as shown in Figure 1. At $r \approx \sqrt{3/\Lambda}$ we have a coordinate singularity in the metric (actually the de Sitter horizon) just as there is a coordinate singularity at $r\approx 2m$ (the Schwarzschild horizon). But these singularities need not concern us here, since the region of interest lies in between.

Figure 1 is a useful picture to have in mind. Here we have plotted, first, on a flat $r,\phi$ plane $\Sigma^1$, the graph of a typical photon orbit $\mathcal{L}^1$ as given by the $r,\phi$ relativistic orbit equation, from which, by general agreement, $\Lambda$ is absent. Exactly the same $r,\phi$ relation holds on the Flamm paraboloid $\Sigma^3$ and on the SdS sphere $\Sigma^2$. Hence the photon orbits $\mathcal{L}^3$ in the Schwarzschild spacetime, and $\mathcal{L}^2$ in the SdS spacetime, correspond to the vertical projections of $\mathcal{L}^1$
 onto the surfaces $\Sigma^3$ and $\Sigma^2$, respectively. In the case of Schwarzschild, where $\Sigma^3$ becomes flat at infinity, the asymptotes of $\mathcal{L}^3$ are identical to those of $\mathcal{L}^1$. So the total deflection angle is the same as that of $\mathcal{L}^1$ in the flat space $\Sigma^1$, and indeed it is usually so evaluated.

\begin{figure}
\begin{center}
\includegraphics[width=3.5in,height=2.5in,angle=0]{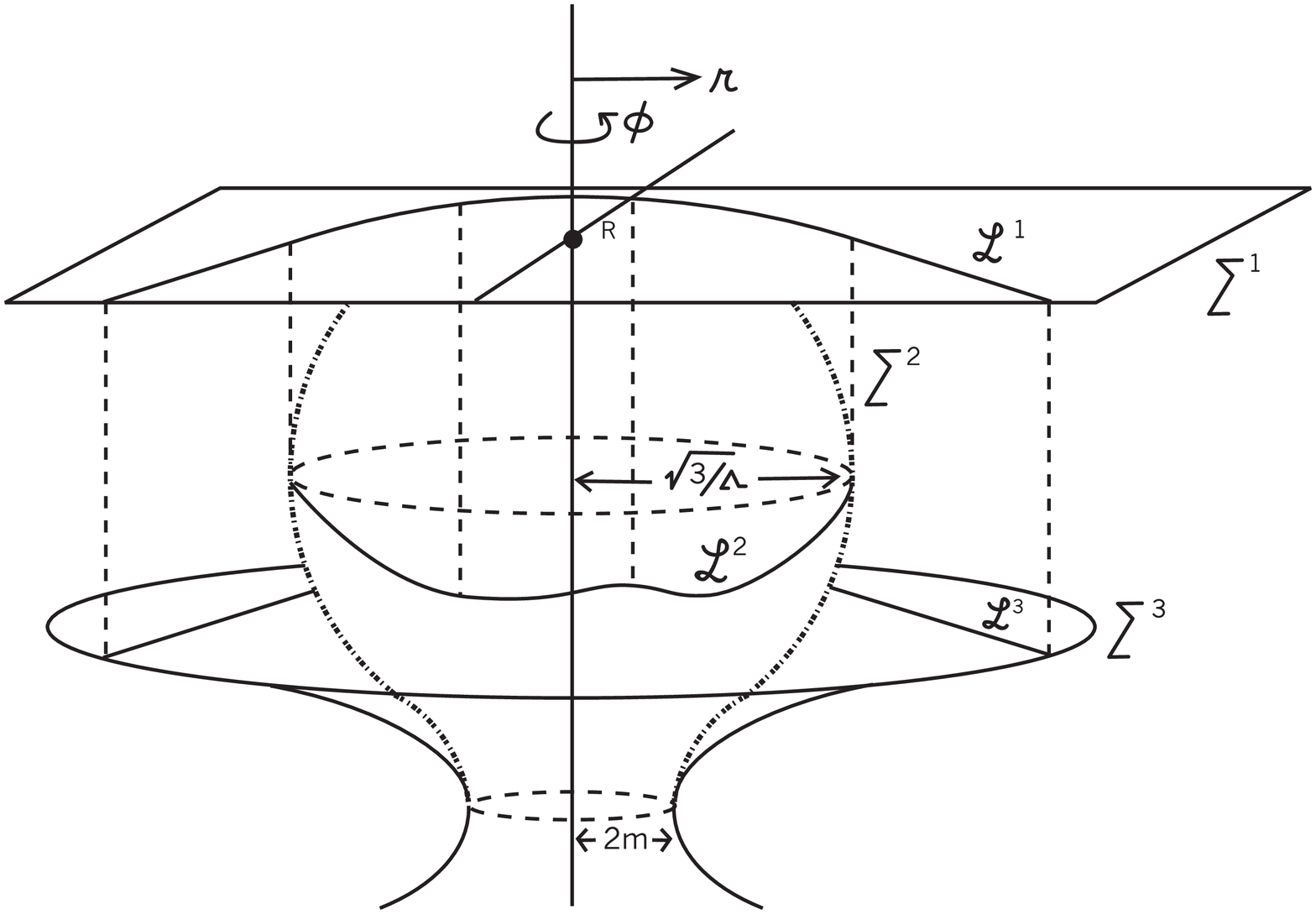}
\caption{\label{fig:figure1} 
Schwarzschild and Schwarzschild-de Sitter geometries.  $\Sigma^3$ is the Flamm paraboloid representation of a central coordinate plane in Schwarzschild;  $\Sigma^2$ is the corresponding surface in Schwarzschild-de Sitter;  $\Sigma^1$ is an auxiliary plane with an $r,\phi$ graph, $\mathcal{L}^1$, of the orbit equation (\ref{eq:solution}).  The curves $\mathcal{L}^2$ and $\mathcal{L}^3$ are the vertical projections of $\mathcal{L}^1$ onto $\Sigma^2$ and $\Sigma^3$, and represent the true spatial curvature of the orbits.
 }
\end{center}
\end{figure}

But for SdS spacetime the situation is more complicated. SdS spacetime in its static representation does {\it not} become flat at infinity. As $r$ approaches the value $\sqrt{3/\Lambda}$, $\mathcal{L}^2$ "climbs up" the SdS sphere to the horizon, and angle measurements differ severely from the corresponding ones in $\Sigma^1$ and $\Sigma^3$. We therefore investigate next how $\Lambda$ affects deflection {\it measurements} in the finite region between the two horizons. 

As is shown in many text books (see for example Eq. (14.24) in \cite{Rindler} with $h=0$, as justified before Eq. (11.62) there), the orbital equation for light in SdS spacetime is 
\begin{equation}
\frac{d^2u}{d\phi^2}+u=3 m u^2, \,\,\,\, (u \equiv 1/r),
\label{eq:ODE}
\end{equation} 
without approximation and in spite of the presence of $\Lambda$ in the SdS metric. This equation is the same as, e.g., Eq. (17) in \cite{Islam} and Eq. (22) in \cite{Freire}.

\begin{figure}
\begin{center}
\includegraphics[width=3.5in,height=1.5in,angle=0]{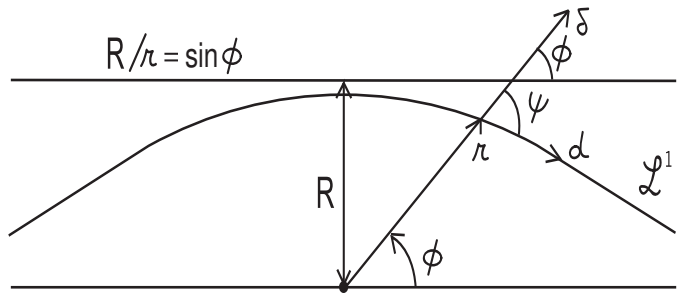}
\caption{\label{fig:figure2} 
The orbital map.  This is a plane graph of the orbit equation (\ref{eq:solution}) and coincides with $\Sigma^1$ in Figure 1. The one-sided deflection angle is $\psi-\phi\equiv\epsilon$.
}
\end{center}
\end{figure}

The orbit that is usually discussed is a small perturbation of the undeflected straight line in flat space
\begin{equation}
r \, sin(\phi)= R
\label{eq:straightline}
\end{equation}
(see Figure 2). This "first approximation" to Eq. (\ref{eq:ODE}) is then substituted into the comparatively small relativistic correction term $3mu^2$, and the resulting linear equation for $u$ solved in the usual way. Thus one obtains (Eq. (11.64) in \cite{Rindler}) 
\begin{equation}
\frac{1}{r}=u=\frac{sin(\phi)}{R}+\frac{3m}{2R^2}\Big{(}1+\frac{1}{3}cos(2\phi)\Big{)}.
\label{eq:solution}
\end{equation}
It is this orbit that we shall take as $\mathcal{L}^1$ in Figure 1 and as the relevant $r,\phi$ relation in both Schwarzschild and SdS spacetime. Its parameter $R$ is related to the physically meaningful area distance $r_{0}$ of closest approach by 
\begin{equation}
\frac{1}{r_{0}}=\frac{1}{R}+\frac{m}{R^2}.
\label{eq:r_0R}
\end{equation}
Other authors, see for example \cite{Wald1984,MTW}, used the impact parameter $b$ to discuss the bending of light in Schwarzschild spacetime,  but the SdS spacetime is not asymptotically flat and one need to define other parameters and constants of motion such as $R$.
 
In Schwarzschild spacetime, for the asymptote of the orbit, we can let $r \longrightarrow \infty$ in (\ref{eq:solution}) and, correspondingly, $\phi \longrightarrow \phi_{\infty}$ (small). Thus we find at once that $\phi_{\infty}=-2m/R$. So the total deflection, defined as the angle between the asymptotes (both in $\Sigma^1$ and $\Sigma^3$), is $4m/R$, as usual. But in SdS spacetime,  $r \longrightarrow \infty$ makes no sense. What we can measure here are the various angles $\psi$ that the photon orbit makes with successive coordinate planes $\phi=const$. (See Figure 2).

Some interesting points were discussed in \cite{Lake2002} and it has been argued there that one should consider the turning point $r_0$ (point of closest approach) and not the constant of motion $b$ in the null geodesic differential equation. Then since $r_0$ is not affected by $\Lambda$, and by virtue of the null geodesic equation, $\Lambda$ should not contribute to the bending of light. As explained in the introduction, this and other arguments were based on the null geodesic differential equation, however, as we demonstrate in the next section, the contribution of $\Lambda$ to the bending angle comes from the spacetime metric itself \cite{metric}, independently from what one will use for the parameterization of the null geodesic differential equation (an independent argument is also discussed in the note \cite{WeylFocusing}).

\section{results and discussion}

In order to calculate the bending angle, we use the invariant formula for the cosine of the angle between two coordinate directions $d$ and $\delta$ as shown in Figure 2 (whose proof is immediate by going to locally Euclidean coordinates):
\begin{equation}
cos (\psi)= {\frac{g_{ij} d{^i} \delta^{j}}{(g_{ij} d^{i} d^{j})^{1/2}(g_{ij}\delta^{i} \delta^{j})^{1/2}}}.
\label{eq:cosine}
\end{equation}
For our purpose the relevant $g_{ij}$ is the 2-metric of $\Sigma^2$, namely (\ref{eq:SdS2metric}). Then 
\begin{equation}
g_{11}=\alpha(r)^{-1}=\Big{(}1-\frac{2m}{r}-\frac{\Lambda r^2}{3}\Big{)}^{-1}, \,\,\,\,\,g_{22}=r^2.
\end{equation}
Next, we differentiate (\ref{eq:solution}) and multiply by $r^2$, to find
\begin{equation}
\frac{dr}{d\phi}=\frac{m r^2}{R^2} sin (2 \phi)-\frac{r^2}{R} cos (\phi)\equiv A(r,\phi).
\label{eq:A}
\end{equation}
Then, if we call the direction of the orbit $d$ and that of the coordinate line $\phi=const$ $\delta$, we have 
\begin{eqnarray}
d      & = & (dr,\, d\phi) = (A,\,\, 1) \, d\phi   \,\,\,\,\,\,  (d\phi < 0) \nonumber \\
\delta & = & (\delta r,\,\, 0)\,\,\,\, = (1,\,\, 0) \, \delta r.
\end{eqnarray}
Substituted into (\ref{eq:cosine}), these values yield 
\begin{equation}
cos(\psi)=\frac{|A|}{(A^2+\alpha(r)r^2)^{1/2}}
\end{equation}
or more conveniently
\begin{equation}
tan(\psi)=(sec^2(\psi)-1)^{1/2}=\frac{\alpha(r)^{1/2}r}{|A|}.
\label{eq:tan}
\end{equation}
The one-sided bending angle is given by $\epsilon=\psi-\phi$. 

Let us calculate $\epsilon=\psi=\psi_0$ when $\phi=0$. By (\ref{eq:solution}), this occurs when $r=R^2/2m$ and consequently $|A|=R^3/4m^2$. Eq. (\ref{eq:tan}) then yields for the (small) angle $\psi_0$:
\begin{equation}
\psi_0 \approx \frac{2m}{R} \Big{(}1-\frac{2m^2}{R^2}-\frac{\Lambda R^4}{24 m^2} \Big{)}.
\label{eq:result1}
\end{equation}

This is the formula of most astrophysical significance. Twice $\psi_0$ is the total bending of a light ray by a massive object, if both source and observer are "far" from that object. In Schwarzschild space we would simply let $r \longrightarrow \infty$ in (\ref{eq:tan}) to get this angle. In SdS space, on the other hand, $r$ cannot exceed its horizon value $\sqrt{3/\Lambda}$, and even that value is unrealistic. The only other intrinsically characterized $r$ value for this purpose is that at $\phi=0$. As Eq.~(\ref{eq:result1}) shows, at that point the classical Einstein one-sided bending angle $2m/R$ has already been reached, to first order, when $\Lambda=0$ (we recall here that $R$ is related to $r_0$ by (\ref{eq:r_0R}) and, to first order, the Einstein angle has the same expression in term of $R$ or $r_0$). And beyond that point we find ourselves in the very extensive, essentially flat, region of transition between Schwarzschild and de Sitter geometry, in which no further significant bending takes place (note that this holds only for particular values of m, R and $\Lambda$). It is in that region that both the source and the observer may be assumed to be situated, and where the observer measures the physical bending angle $2\psi_0$ directly as the angle between the apparent and the undisturbed position of the source. Note, from (\ref{eq:result1}), that a positive $\Lambda$ diminishes the bending angle, as expected.

For completness' sake it is of interest also to look at bending angles occurring at $\phi$-values other than zero, and their connection with observations. As an example, we calculate $\psi$ when $\phi=45^\circ$. To sufficient accuracy, (\ref{eq:solution}) is then satisfied by $r=\sqrt{2}R$. With that, and assuming $m/R$ and $\Lambda R^2$ to be small, we find successively from (\ref{eq:A}), (\ref{eq:alpha}), and (\ref{eq:tan}):
\begin{eqnarray}
A &=&2m-\sqrt{2}R= -\sqrt{2}R\Big{(}1-\frac{\sqrt{2}m}{R}\Big{)} \nonumber \\
\alpha^{1/2}&=&\Big{(}1-\frac{2m}{\sqrt{2}R}-\frac{2}{3}\Lambda R^2\Big{)}^{1/2}=1-\frac{m}{\sqrt{2}R}-\frac{\Lambda R^2}{3} \nonumber \\
tan(\psi)&=&1+\frac{m}{\sqrt{2}R}-\frac{\Lambda R^2}{3}.
\label{eq:result2}
\end{eqnarray}
The actual (small) bending angle at $\phi=45^\circ$ is $\epsilon=\psi-\phi$. For this, we find, from (\ref{eq:result2}), 
\begin{equation}
\epsilon=\frac{m}{2\sqrt{2}R}-\frac{\Lambda R^2}{6},
\label{eq:result3}
\end{equation}
where we used $\epsilon=\psi-\phi\approx tan(\psi-\phi)=(tan(\psi)-1)/(1+tan(\psi))$ and $tan(\phi)=1$. Once again there is the expected negative contribution from $\Lambda$.

The observational situation in the general case is as follows. The observer is at the intersection of the ray coming from the distant source and a radius having coordinate direction $\phi$. Clearly such an observer can measure $\psi$ directly: it is the visually determined angle from the center of the lens to the apparent position of the star being lensed.  In principle, $\phi$ is also measurable, but it requires the input of a second observer far away who has determined the total bending angle $2\psi_0$. Then $\phi=\beta+\psi_0$, where $\beta$ is the angle from the center of the lens to the undisturbed position of the star. We provided here a brief prescription for observations, however, a full study on how to put our results into an observational context is pursued and will be presented in a follow up paper. We will also include there second order perturbations to the geodesic equation in order to match the higher order terms in equation (\ref{eq:result1}).

Of course, the contribution of $\Lambda$ to the bending of light is very small. We know from cosmology that $\Lambda$ is of the order of $10^{-56} cm^{-2}$. With that, the ratio of the two terms on the RHS of (\ref{eq:result1}), in the case of a ray grazing the limb of the Sun, is $10^{28}:1$! In this respect, though it is also hopelessly small, the contribution of $\Lambda$ to the advance of the perihelion of Mercury (see Eq. (14.25) in \cite{Rindler}) is superior: it could be as much as $10^{-15}$ of the total. However, one would  expect this to be different at very large scales (e.g. clusters and superclusters of galaxies) and this will be explored in detail in a follow up paper.  

In conclusion, the present paper provides a long overdue correction to previous
 works on whether or not the cosmological constant $\Lambda$ affects the bending of light by a concentrated spherically symmetric mass. We showed that when the geometry of the Schwarzschild-de Sitter spacetime is taken into account,  $\Lambda$ indeed contributes to the light-bending. 

\acknowledgements 
M.I. acknowledges partial support from the Hoblitzelle Foundation and a Clark award at UTD.


\begin{thebibliography}{}
\bibitem{metric}It is the metric that measures the distances in the spacetime.
\end{thebibliography}
\end{document}